\documentstyle[aps,floats,epsf,psfig,twocolumn]{revtex}
\begin{document}
\tighten
\draft

\title{Stripe Orientation in an Anisotropic $t$-$J$ Model}

\author{Arno P.~Kampf$^1$\thanks{kampfa@physik.uni-augsburg.de},
Douglas J.~Scalapino$^2$\thanks{djs@vulcan.physics.ucsb.edu}, and
Steven R.~White$^3$\thanks{srwhite@uci.edu}}
\vskip0.3cm
\address{$^1$ Institut f\"ur Physik, Theoretische Physik III, Elektronische
Korrelationen und Magnetismus, \\ Universit\"at Augsburg, 86135 Augsburg,
Germany}
\address{$^2$ Department of Physics,
         University of California,
         Santa Barbara, CA 93106-9530}
\address{$^3$ Department of Physics,
         University of California,
         Irvine, CA 92697}
\address{~
\parbox{14cm}{\rm 
\medskip
The tilt pattern of the $CuO_6$ octahedra in the LTT phase of the cuprate
superconductors leads to planar anisotropies for the exchange coupling and
hopping integrals.  Here, we show that these anisotropies provide a
possible structural mechanism for the orientation of stripes.  A
$t_x$-$t_y$-$J_x$-$J_y$ model thus serves as an effective Hamiltonian to
describe stripe formation and orientation in LTT-phase cuprates.
}}
\maketitle

\narrowtext

Early Hartree-Fock calculations \cite{ZG89} found evidence for domain-wall
formation in doped 2D Hubbard and $t$-$J$ models.  In these calculations
the domain walls contained one hole per unit cell and separated $\pi$-phase
shifted antiferromagnetic (AF) regions.  Subsequent
density-matrix-renormalization-group (DMRG) calculations \cite{WSxx} also
found hole-domain walls separating $\pi$-phase shifted AF regions, but in
these calculations the linear filling of the horizontal (or vertical)
domain walls corresponded to one hole per two unit cells of the wall.  In
these calculations, domain-wall formation originates as a compromise
in the inherent competition between the kinetic and exchange energies
which arises when holes are added to a Mott antiferromagnetic insulator.
In the parameter regime where horizontal or vertical stripes formed, these
four-fold rotationally invarient models did not distinguish between the
two orientations.  Here we wish to discuss a possible electronic mechanism
for stripe orientation.

We are motivated by the structural phase transition \cite{Buc94} of
$La_{1.6-x} Nd_{0.4} Sr_x CuO_4$, in which the system goes from a low
temperature orthorhombic (LTO) to a low temperature tetragonal (LTT) phase
below $\sim 70K$.  Here, as illustrated in Fig.~1a, in the LTT phase the
$CuO_6$ octahedra tilt around an axis oriented along the planar $Cu$-$O$
bonds, say along the $y$-direction.  As a consequence, oxygen atoms on
the tilt axis remain in the plane, but in the perpendicular $x$-direction
a staggered tilting pattern results with oxygen atoms $O_a$ and $O_b$ in
Fig.~1a displaced above or below the $CuO_2$ plane, respectively. The $x$-
and $y$-directions are therefore no longer equivalent in contrast to the
LTO phase, where the tilt axis is rotated by 45 degrees, as shown in Fig.~1b. 

\begin{figure}[t!]
\psfig{file=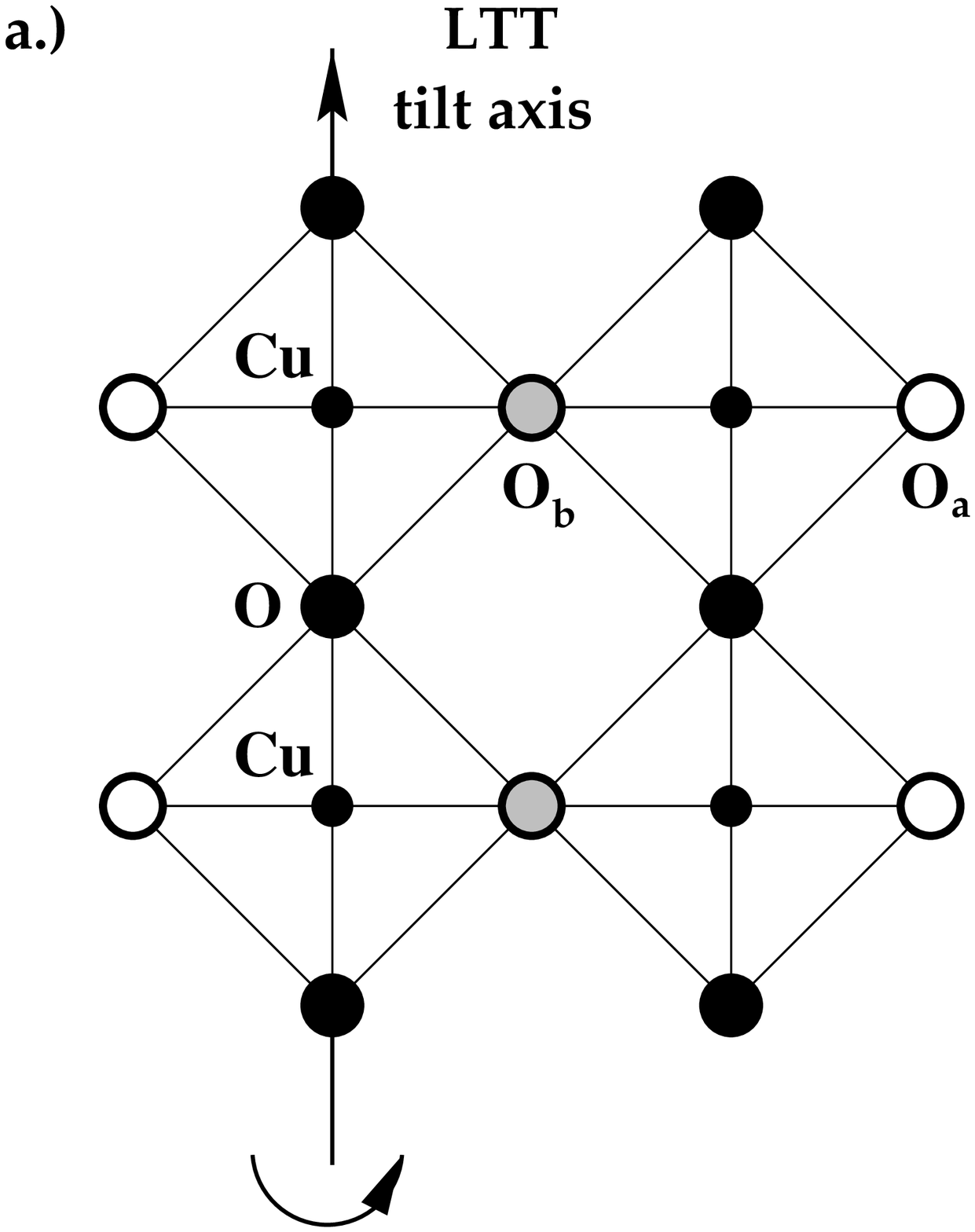,height=9.2cm,width=8.cm,angle=0}
\vskip0.2cm
\psfig{file=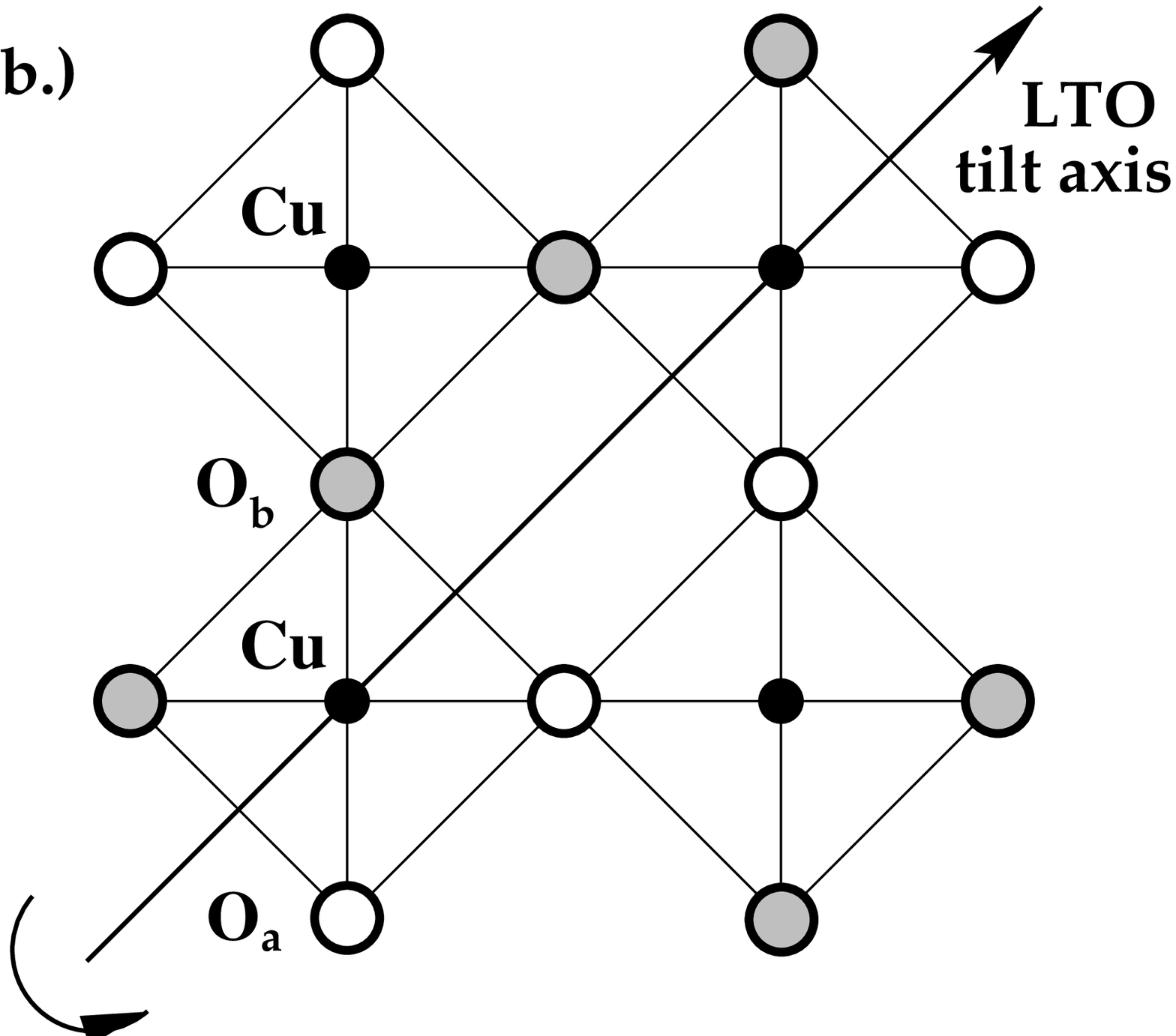,height=7.2cm,width=9.cm,angle=0}
\vskip0.5cm
\caption[]{Planar view of the tilt pattern of the $CuO_6$ octahedra in the
(a) LTT and (b) LTO phase. In (a) oxygen atoms along the vertical bonds
remain in the $CuO_2$ plane while in the perpendicular direction they move
below $(O_b)$ or above $(O_a)$ the plane in a staggered pattern, leading to
a reduction of $t_x$ and $J_x$ relative to $t_y$ and $J_y$.}
\label{fig1}
\end{figure} 

The electronic hopping integrals, and thus the antiferromagnetic
superexchange, in the $CuO_2$ planes depends sensitively on the
$Cu$-$O$-$Cu$ bond angle $\theta$.  In the specific buckling pattern of the
LTT phase, this bond angle is $\theta_y=\pi$ along the tilt axis direction,
but is reduced by twice the octahedral tilt angle $\alpha$ in the
perpendicular direction, i.e. $\theta_x=\pi-2\alpha$. This $x$-$y$
anisotropy for the electronic hopping and superexchange parameters may be
conveniently translated into an anisotropic $t$-$J$ model Hamiltonian
\begin{eqnarray}
H= &-& t_x \sum_{\langle i, i+x\rangle\sigma} \left(c^\dagger_{i+x\sigma}
c_{i\sigma} + h.c.\right) \nonumber\\
&+& J_x \sum_{\langle i,i+x\rangle} \left({\bf S}_{i+x}
\cdot {\bf S}_i- \frac{n_{i+x}n_i}{4}\right)\nonumber\\
   &-& t_y \sum_{\langle i,i+y\rangle\sigma} \left(c^\dagger_{i+y\sigma}
c_{i\sigma} + h.c.\right) \nonumber\\
&+& J_y \sum_{\langle i,i+y\rangle} \left({\bf S}_{i+y}
\cdot {\bf S}_i - \frac{n_{i+y}n_i}{4}\right)\ .
\label{one}
\end{eqnarray}
Here, $\langle i,i+x\rangle$ and $\langle i,i+y\rangle$ denote
nearest-neighbor sites along the $x$- and $y$-directions on a square
lattice, respectively, and doubly-occupied sites are explicitly excluded
from the Hilbert space. 

The magnitude of the anisotropies is easily estimated for typical tilt
angles of $4^\circ$--$5^\circ$ in $La_{1.6-x} Nd_{0.4} Sr_x CuO_4$ with $x$
near 1/8 \cite{Buc94}.  When the tilt axis of the LTT phase is vertical, as
shown in Fig.~1a, we have
\begin{equation}
\frac{t_x}{t_y} \cong |\cos (\pi-2\alpha)| \qquad {\rm and} \qquad
\frac{J_x}{J_y} \cong \cos^2 (\pi-2\alpha)
\label{two}
\end{equation}
It follows that for a tilt angle of order $4^\circ$--$5^\circ$, $\Delta
t/t\sim 1.$--1.5\% and $\Delta J/J\sim 2.$--3.\%. We note that the
direction with the
larger exchange coupling is naturally also the direction with the larger
hopping amplitude.  Choosing $t=500$meV and with the exchange coupling
constant $J=1500K$ of undoped $La_2CuO_4$ these estimates give
$\Delta t= |t_x-t_y| \sim 60K$ and $\Delta J= |J_x - J_y| \sim 40K$. This
rough estimate for the exchange anisotropy agrees with results from
quantum chemistry calculations \cite{Marxx}. 

Given this model Hamiltonian, with $J_y>J_x$ and $t_y>t_x$, one may ask in
which direction stripes are expected to form.  Since $J_y>J_x$, the exchange 
energy is optimized by orienting the domain walls along the $y$-axis so
as to  minimize the number of broken exchange bonds in the direction with
the stronger superexchange.  Now, one might be tempted to argue that since
$t_y>t_x$, this also lowers the kinetic energy of the system. However, 
transverse motion of the domain walls is also known to be important 
\cite{ZG89,NKxx,ZO96}, so that an anisotropy in the hopping with
$t_y>t_x$ can favor a horizonal orientation of the stripes. Because $\Delta
J$ and $\Delta t$ are comparable in magnitude, we analyze the results 
of a DMRG calculation to obtain further insight in this point.

We have used DMRG techniques to study a $9\times 8$ lattice with periodic
boundary conditions in the 8-site $y$-direction and open boundary 
conditions in the 9-site $x$-direction. 
Fig.2a shows a domain which forms when 4 holes are added for an isotropic
Hamiltonian with $J/t=0.35$.  The boundary conditions cause the domain to
form around the middle of this 8-leg cylinder. According to the
Hellman-Feynman theorem, 
\begin{equation}
\frac{\partial \langle H\rangle}{\partial J_x} = \sum_{\langle 
ij\rangle \atop x{\rm -bonds}}
\left\langle \vec S_i \cdot \vec S_j - \frac{n_in_j}{4}\right\rangle
\label{three}
\end{equation}
Therefore, by calculating the change
\begin{eqnarray}
\Delta \left\langle\vec S_i\cdot \vec S_j - \frac{n_in_j}{4}\right\rangle
&=&\left\langle\vec S_i\cdot\vec S_j-\frac{n_in_j}{4}\right\rangle_4\nonumber\\
&-&\left\langle\vec S_i\cdot \vec S_j - \frac{n_in_j}{4}\right\rangle_o
\label{four}
\end{eqnarray}
between the expectation value in the 4-hole ground state and the undoped
ground state for the isotropic case with a given value of $J/t$, we can
determine the variation of the domain wall energy with respect to small
changes in $J_x$ near $J$. The local change for the individual $x$-bonds,
eq.~(\ref{four}), which contribute to $\partial\langle
H\rangle/\partial J_x$ are shown on the horizontal $x$-bonds in Fig.~2b
for $J/t=0.35$. Note that these contributions decrease as one moves away
from the domain wall and we find that
\begin{eqnarray}
\frac{1}{4}\ \frac{\partial}{\partial J_x}\ \left(\langle H\rangle_4 -
\langle H\rangle_o\right) &=& -\frac{1}{4}\ \sum_{\langle
ij\rangle \atop x{\rm -bonds}} \Delta \left\langle\vec S_i\cdot \vec S_j - 
\frac{n_in_j}{4}\right\rangle\nonumber\\
 &=& 1.91/{\rm hole}\ .
\label{five}
\end{eqnarray}

\begin{figure}[t!]
\psfig{file=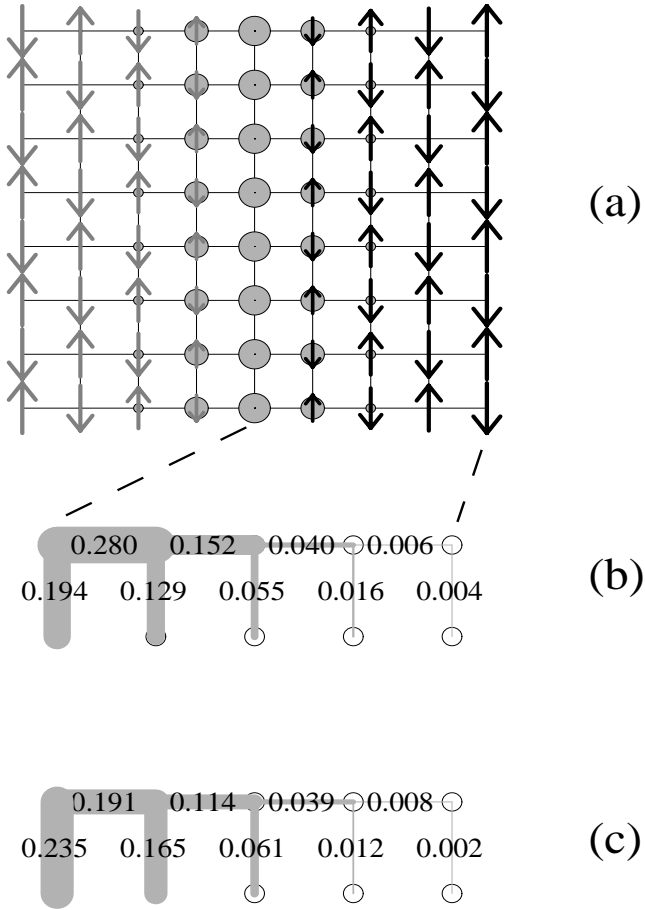,height=12.7cm,width=9.cm,angle=0}
\vskip0.5cm
\caption[]{(a) A $9\times 8$ lattice with $J/t=0.35$ and 4 holes which form a
site-centered domain wall. This lattice has periodic boundary conditions in
the $y$-direction and open ends in the $x$-direction where a weak, 
$\pi$-phase shifted staggered magnetic field $h= \pm0.1t$, is applied at the 
open ends. The diameter of the circles indicate the hole density and the 
length of the spins, the spin magnitude. The length of the arrows on the left
and right sides corresponds to $\langle S_z\rangle=0.37$. The hole density, 
which is proportional to the diameter of the gray circles, is 
$\langle n_h\rangle = 0.155$ along the center of the stripe.
(b) The change in $\langle(\vec S_i\cdot \vec S_j
- \frac{1}{4} n_i n_j)\rangle$ between the 4-hole ground state with the 
domain wall and the
undoped Heisenberg lattice. In the undoped case the weak applied
staggered end-field was periodic. The value 0.194 corresponds to the axis
of the domain wall, and the values are symmetric about this axis. (c) The
kinetic energy $-\sum_s \langle(c^\dagger_{is}c_{js}+h.c.)\rangle$ 
in units of $t$ of the
4-hole system.}
\label{fig2}
\end{figure}

In a similar manner we find that the variation of the domain wall energy
with the exchange energy $J_y$ parallel to the wall gives
\begin{eqnarray}
\frac{1}{4}\ \frac{\partial}{\partial J_y}\, \left(\langle H\rangle_4
- \langle H\rangle_o\right)
&=& - \frac{1}{4}\ \sum_{\langle ij\rangle \atop y{\rm -bonds}} \Delta 
\left\langle\vec S_i\cdot \vec S_j - \frac{n_in_j}{4}\right\rangle\nonumber\\
&=& 1.20/{\rm hole}\ .
\label{six}
\end{eqnarray}
Continuing with the kinetic energy terms,
\begin{eqnarray}
\frac{1}{4}\ \frac{\partial}{\partial t_x}\ \langle H\rangle_4 &=& -
\frac{1}{4} \sum_{\langle ij\rangle s \atop x{\rm -bonds}} \left\langle
c^\dagger_{is} c_{js} + c^\dagger_{js} c_{is}\right\rangle\nonumber\\
 &=& - 1.43/{\rm hole}
\label{seven}
\end{eqnarray}
and
\begin{eqnarray}
\frac{1}{4}\ \frac{\partial}{\partial t_y}\ \langle H\rangle_4 &=& 
-\frac{1}{4} \sum_{\langle ij\rangle s \atop y{\rm -bonds}}\left\langle 
c^\dagger_{is} c_{js} + c^\dagger_{js} c_{is}\right\rangle\nonumber\\ 
&=& -
1.43/{\rm hole}\ .
\label{eight}
\end{eqnarray}
The local kinetic energy of the 4-hole systems associated with the domain
wall are listed in Fig.~2c.

Now, if the tilt axis of the LTT structure runs along the $y$-axis so that
it is parallel to the domain wall, then $J_y=J$, $J_x=J-\Delta J$, $t_y=t$,
and $t_x=t-\Delta t$. In this case, the shift in energy per hole of the
domain wall due to the small anisotropy is
\begin{equation}
\Delta E_\| = -1.9 \Delta J+ 1.41 \Delta t\ .
\label{nine}
\end{equation}
Alternatively, if the LTT tilt axis runs along the $x$-axis, perpendicular
to the domain wall, the shift in energy per hole is
\begin{equation}
\Delta E_\perp = - 1.20 \Delta J+1.43 \Delta t\ .
\label{ten}
\end{equation}
Therefore, if the domain wall is oriented parallel to the LTT tilt axis,
there is a net energy reduction (relative to an orientation perpendicular
to the tilt axes) of
\begin{equation}
\Delta E= 0.7 \Delta J- 0.02 \Delta t \simeq 25K/{\rm hole}\ .
\label{eleven}
\end{equation}
For a section of domain wall containing 4 holes, this would be 100K. The
extensive nature of this energy favors alignment of the domain wall with
the LTT tilt axis.

Thus we conclude that the $\Delta J$ anisotropy dominates and favors
orienting the stripes along the direction of the tilt axis of the LTT
phase.  This is the same orientation as suggested from the ``structural
corrugation'' driven orientation mechanism originally set forth by
Tranquada et.~al. \cite{Tran95}. Here, we have simply looked at a
particular model in which the corrugation manifests itself by giving rise
to an anisotropic $t$-$J$ model.

\acknowledgments

APK would like to acknowledge support from the Deutsche
Forschungsgemeinschaft through SFB 484. DJS and SRW would like to
acknowledge support from the US Department of Energy under Grant
No.~DE-FG03-85ER45197.

\end{document}